# Miniaturized backward coupler realized by the circuit-based magnetic hyperbolic waveguide

Zhiwei Guo[*], Juan Song, Haitao Jiang, and Hong Chen[†]

*Key Laboratory of Advanced Micro-structure Materials, MOE, School of Physics Science and Engineering, Tongji University,*

*Shanghai 200092, China*

**Abstract**

Planar waveguides can limit the transmission of electromagnetic waves in a specific direction and have a wide range of applications in filters, sensors, and energy-transfer devices. However, given the increasing demand for planar integrated photonics, new waveguides are required with excellent characteristics such as more functionality, greater efficiency, and smaller size. In this work, we report the experimental results for a subwavelength planar microwave-regime waveguide fabricated from circuit-based magnetic hyperbolic metamaterial (HMM) and develop the associated theory. HMM is a special type of anisotropic metamaterial whose isofrequency contour (IFC) takes the form of an open hyperboloid. Because of the open-dispersion IFCs, HMMs support propagating high-$k$ modes with large effective refractive indices, which allows planar hyperbolic waveguides to be miniaturized. In particular, as opposed to the traditional dielectric slab waveguide, the group velocity of the hyperbolic guided modes is negative—a characteristic that can be exploited to realize the backward propagation of electromagnetic waves. Furthermore, we study the abnormal photonic spin Hall effect and experimentally demonstrate an ultracompact backward coupler based on hyperbolic waveguides. The experimental results are consistent with numerical simulations. This work not only reveals the significant potential of circuit-based platforms for the experimental study of the propagation and coupling of guided modes but also promotes the use of HMMs in the microwave regime for numerous integrated functional devices.

Corresponding authors:

[*] 2014guozhiwei@tongji.edu.cn

[†] hongchen@tongji.edu.cn

# I. INTRODUCTION

The manipulation of electromagnetic waves depends strongly on the isofrequency contour (IFC) of the medium. By modifying the shape of the momentum-space IFC in a given medium, one can flexibly control the propagation, coupling, and radiation characteristics of electromagnetic waves [1–6]. In particular, when the IFC changes from a closed ellipsoid to an open hyperboloid, the interaction between electromagnetic waves and matter is strongly modified by the topological properties of the structure changes [7–10]. For normal anisotropic materials, the IFC is a closed ellipsoid, and the wave vector magnitude is limited by the boundary of the closed IFC. However, the IFC of hyperbolic metamaterials (HMMs) is an open hyperboloid, which allows propagating waves with large wave vectors [11, 12]. The high-$k$ modes in HMMs significantly enhance the photonic density of states and can be used to improve the rate of spontaneous emission [8, 13, 14]. The unique characteristics of HMMs can also be used to overcome many physical limitations of normal materials. For example, the long-range dipole-dipole interaction realized by HMMs surpasses the distance limitation of near-field coupling and can be used to realize long-range energy transfer [15–18] or enhanced electromagnetically induced transparency [19], and super-resolution imaging realized by HMMs overcomes the diffraction limit of traditional imaging systems and can be used to construct hyper-lenses [20–23]. Furthermore, the high-quality-factor microcavity made possible by HMMs breaks the size limitation of traditional microcavities and can be used to fabricate miniaturized lasers and filters [24–26]. Recently, other interesting research topics and applications have been enabled by HMMs, including optical pulling forces [27, 28], the giant Unruh effect [29], topological and non-Hermitian systems [30–32], high-sensitivity sensors [33–35], fingerprinting [36], and wavefront manipulation [37]. Therefore, HMMs provide a new material system and physical mechanism that allows us to control the transmission of electromagnetic waves.

Developing techniques to restrict electromagnetic waves and make them transmit along a specific channel is highly significant for optical integration. By using the mechanism of total internal reflection, HMMs can be used to fabricate waveguides with special physical properties [38–45]. To date, electric HMMs (the principal components of its permittivity tensor have opposite signs) constructed by metal/dielectric multilayer structures have been widely used in the design of hyperbolic waveguides and exploit some novel physical effects such as optical field enhancement [38], rainbow trapping [39], slow light [42], and long-range transmission [43]. In these structures, HMMs mainly serve as waveguide reflection walls, which have certain advantages. On the one hand, compared with metal walls, less loss occurs with the HMM wall. On the other hand, compared with a dielectric wall, a HMM wall provides greater confinement strength. However, hyperbolic waveguides are limited to transverse-magnetic–polarized (TM-polarized) waves (i.e., the electric field is parallel to the optical axis), and the multilayer structure makes it difficult to realize photon integration and measure the field distributions because of the three-dimensional structure. Moreover, using the HMM as the core layer of the waveguide may simplify the hyperbolic waveguide and make it more practical.

Here we propose, design, and fabricate planar magnetic hyperbolic waveguides by using the circuit-based magnetic HMM (the principal components of its permeability tensor have opposite signs). It should be noted that the magnetic HMM here is not the gyromagnetic material and has no magneto-optical response. These waveguides are based on two-dimensional (2D) transmission lines (TLs). On the one hand, magnetic HMMs have attracted increasing attention because of their ability to manipulate transverse-electric−polarized (TE-polarized) waves (i.e., the electric field is perpendicular to the optical axis) [46, 47]. On the other hand, similar to metasurfaces [48–55] and natural 2D materials [56–59] with hyperbolic dispersion, circuit-based HMMs are planarized, which facilitates the measurement of the spatial distribution of near-field modes. Importantly, active control may be implemented in circuit-based HMMs by applying an external voltage [60–63]. Therefore, circuit-based magnetic hyperbolic waveguides provide an important research platform for exploring near-field control. In circuit-based hyperbolic waveguides, HMM is used as the core layer, whereas normal double-positive (DPS) material is used as the cladding. The serious mismatch between high-$k$ modes in HMM and propagating modes in the surrounding medium leads to strong total internal reflection. Therefore, HMMs with high-$k$ modes can be used to design high-performance waveguides with high confinement. In this case, electromagnetic waves are well transmitted in the hyperbolic core but attenuate exponentially in the DPS cladding. By comparing the dispersion relation of the normal waveguide with that of the hyperbolic waveguide, we find that the hyperbolic waveguide can support backward-guided modes. By using this characteristic, we can study the anomalous photonic spin Hall effect (PSHE) and design a miniaturized backward coupler based on hyperbolic waveguides. Moreover, the results of research into near-field control of electromagnetic waves in circuit-based hyperbolic waveguides provide broad guidelines for developing directional antennas and miniaturized filters.

This work is organized as follows: Section II covers the design of the circuit-based magnetic hyperbolic waveguide and the associated physical properties. Section III investigates the coupling mechanism of hyperbolic waveguides and, in particular, the related applications of abnormal PSHE and ultracompact backward coupling in the hyperbolic waveguide. Finally, Sec. IV summarizes and concludes this work.

## II. DESIGN OF CIRCUIT-BASED HYPERBOLIC WAVEGUIDE AND GUIDED MODES WITH BACKWARD-WAVE CHARACTERISTICS

Various circuit-based metamaterials can be conveniently implemented by using the TL system loaded with lumped circuit elements, including two kinds of single-negative media, double-negative media, zero-index media, etc. [64, 65]. A series of important applications of circuit-based metamaterials exist in the microwave regime, such as filters [66] and frequency mixers [67]. Recently, circuit-based HMMs and the associated topological transition have been studied [68–70]. Here, we expand upon a 2D TL system to design a hyperbolic waveguide that uses circuit-based magnetic HMMs (see Fig. 1). The hyperbolic waveguide is composed of a core layer of HMM and two cladding layers consisting of a DPS medium. The DPS and HMM regions are colored green and red, respectively, to facilitate viewing. In particular, the HMM was realized by loading lumped capacitors ($C$ = 5 pF) in the $x$ direction, whereas the DPS medium corresponds to the structure without lumped capacitors. The unit structures of the HMM and DPS media are enlarged on the right side of Fig. 1. The entire structure was constructed on a commercial printed circuit board F4B, which has a relative permittivity and thickness of $\varepsilon_r = 2.2$, and $h$ = 1.6 mm, respectively. The loss tangent of the F4B board is $\tan\delta = 0.0079$. The width of the metal microstrip is $w = 2.8$ mm, and the length of a unit cell is $p$ = 12 mm. The mode width of the hyperbolic waveguide is $d + 2\delta$, where $d$ and $\delta$ denote the width of the HMM core and the penetration depth on the side of the DPS cladding, respectively.

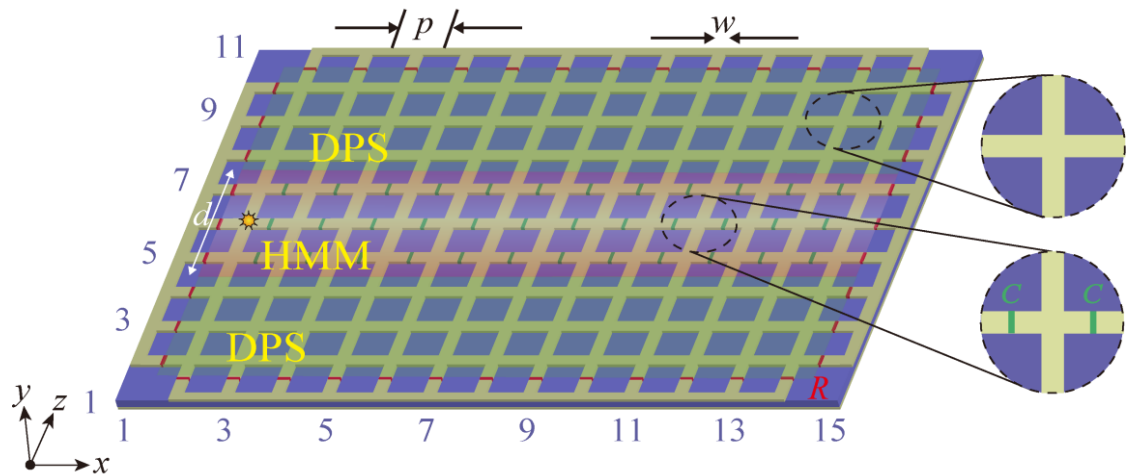

FIG. 1. Schematic of TL-based planar waveguide with HMM core. The hyperbolic layer is sandwiched between two DPS layers. Here, $p$ = 12 mm and $w$ = 2.8 mm. The width of the HMM core is $d$. Unit structures of the HMM and DPS media are enlarged to the right of the main figure.

Figure 2(a) shows a simple, effective circuit model for a circuit-based hyperbolic waveguide. In this circuit model, the two interfaces between the HMM and DPS media are marked by dashed gray lines. The structure factor of this structure is $g = Z_0/\eta_{\text{eff}} \approx 0.255$, where $Z_0$ and $\eta_{\text{eff}}$ denote the characteristic impedance and effective wave impedance, respectively [64, 65]. In the long-wave approximation, the effective electromagnetic parameters of the system when the capacitors are loaded on the TL in the $x$ direction are [68–70]

$$\varepsilon = 2C_0 g/\varepsilon_0, \quad \mu_x = \frac{L_0}{g\mu_0}, \quad \mu_z = \frac{L_0}{g\mu_0} - \frac{1}{\omega^2 Cdg\mu_0}, \tag{1}$$

where $\varepsilon_0$ and $\mu_0$ are the vacuum permittivity and permeability, respectively, $\omega$ is the angular frequency, and $C_0$ and $L_0$ denote the capacitance and inductance per unit length, respectively [68–70]. The electromagnetic parameters calculated by Eq. (1) are illustrated in Fig. 2(b) as a function of frequency. With increasing frequency, $\mu_z$ changes from negative to positive, whereas $\varepsilon \approx 3.63$ and $\mu_x = 1$ remain constant. When the frequency is 1.14 GHz, $\mu_x = 0$, which corresponds to the topological phase transition point from the open hyperbola to the closed ellipse. To show this topological phase transition, we take one frequencies as examples: $f = 0.85$ GHz, which is shown by yellow star in Fig. 2(b). The dispersion relation of the circuit-based metamaterials is [68–70]

$$\frac{k_x^2}{\varepsilon\mu_z} + \frac{k_z^2}{\varepsilon\mu_x} = \left(\frac{\omega}{c}\right)^2, \tag{2}$$

where $k_x$ and $k_z$ are the $x$ and $z$ components of the wave vector, respectively, and $c$ is the speed of light in vacuum. Figures 2(c) shows the measured electric field distribution of the circuit-based HMM based on the near-field detection method, in which the sample is excited by a point source at the center of the structure (see more details in supplementary materials). The corresponding IFC obtained as the Fourier spectrum is shown in Fig. 2(d). The dashed yellow line is the numerically calculated results based on Eq. (2).

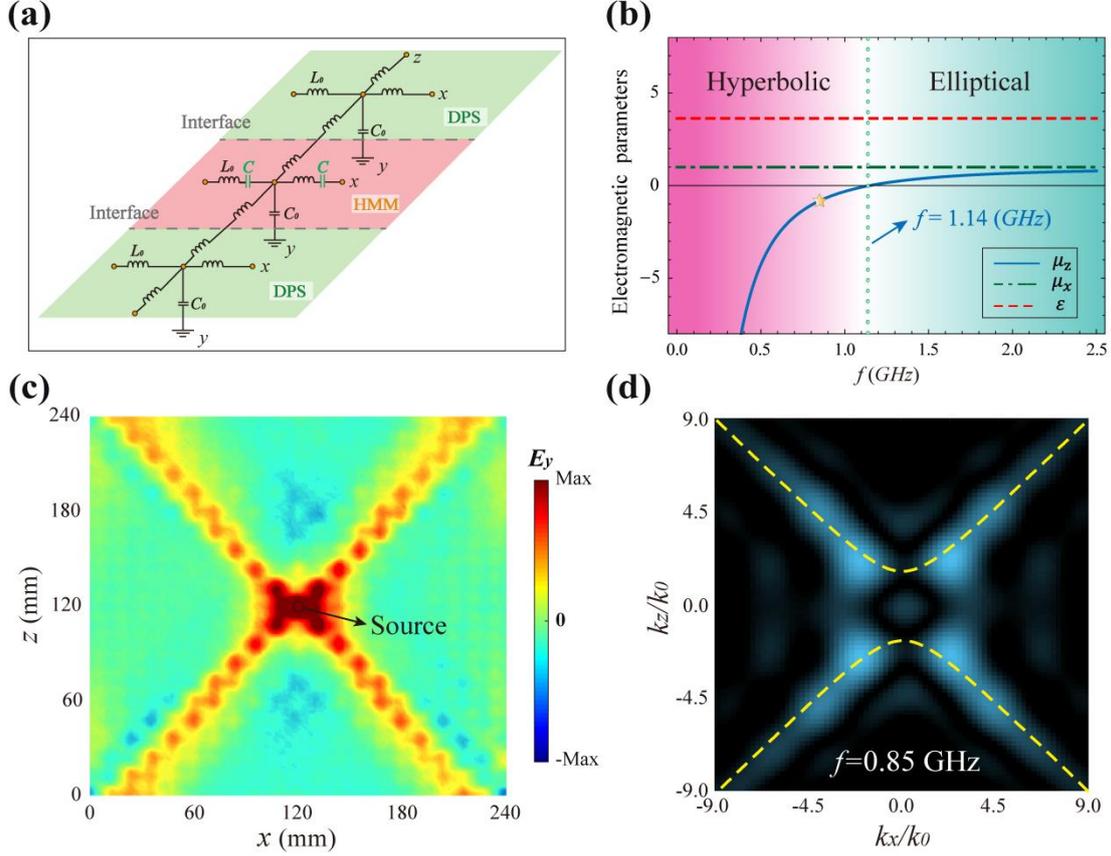

FIG. 2. (a) 2D circuit models of HMM and DPS media. (b) Effective electromagnetic parameters based on TLs when capacitors are loaded in the $x$ direction. The topological transition of the IFC from hyperbolic to elliptical is labeled by the gradient color from pink to green. (c) Measured electric field distribution of the circuit-based HMM at $f = 0.85$ GHz. (d) The corresponding IFC of circuit-based HMM at $f = 0.85$ GHz. The yellow dashed curve shows the analytical dispersion.

Based on the effective electromagnetic parameters in Fig. 2(b), we calculate the dispersion relation of guided modes in the circuit-based hyperbolic waveguide. Since the tangential components of the electric and magnetic fields should be continuous at the boundary of the HMM and DPS media ($\varepsilon_D \approx 3.63$, $\mu_D = 1$), the dispersion relation of the lowest-order TE-polarized guided modes can be deduced from the characteristic equation [38–45]

$$k_{zH}d = 2\arctan\left[\frac{\mu_x \operatorname{Im}(k_{zD})}{\mu_D k_{zH}}\right], \tag{3}$$

where

$$k_{zH} = \sqrt{\frac{\omega^2}{c^2}\varepsilon\mu_x - k_x^2\frac{\mu_x}{\mu_z}}$$

and

$$k_{zD} = \sqrt{\frac{\omega^2}{c^2}\varepsilon_D\mu_D - k_x^2}$$

denote the wave vector perpendicular to the interface in the HMM core and DPS cladding, respectively, and $k_x$ is the propagation wave vector of the guided mode. Figure 3(a) shows the dispersion relations based on Eq. (3) of the circuit-based hyperbolic waveguides for several HMM core widths $d$. The group velocity is calculated by using $v_g = \partial\omega/\partial k_x$, which gives a negative group velocity for the hyperbolic guided modes. To illustrate the confinement of the guided modes, we calculate the penetration depth $\delta = \operatorname{Im}[2k_{zD}]^{-1}$ on one side of the DPS cladding [see Fig. 3(b)]. A lower frequency correlates with stronger confinement. We choose a HMM core width $d = 12$ mm and a frequency $f = 0.7$ GHz, as shown by the pink star in Fig. 3(a). By using full-wave commercial software (CST Microwave Studio), we show the corresponding electric-field distribution of hyperbolic guided mode, as shown in Fig. 3(c). The results clearly show that the electromagnetic waves are strongly confined in the central hyperbolic core layer. The simulation uses perfect boundary matching, as marked by the rhombus in Fig. 3(c). In particular, the abnormal dispersion relation of hyperbolic guided modes reverses the sign of the phase velocity ($v_p = \omega/k_x$) with respect to that of the group velocity. In fact, the direction of the group velocity of the guided mode excited by the excitation source is always positive, so the anomalous dispersion of the hyperbolic waveguide mode is the opposite of the phase velocity [71, 72]. In Fig. 3(c), the white and orange arrows show the directions of the energy flux (group velocity) and wave vector (phase velocity), respectively.

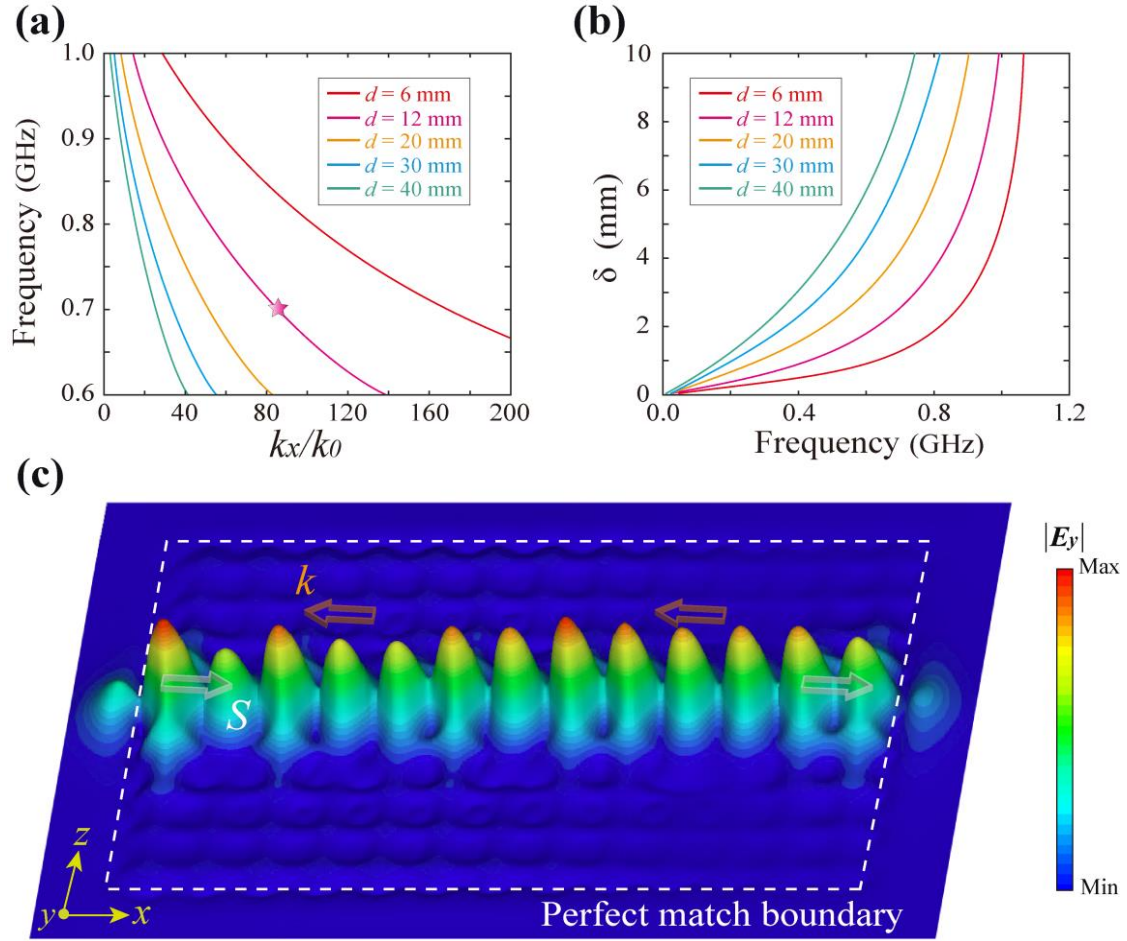

FIG. 3. (a) Dispersion relations of guided modes in hyperbolic waveguides of various core widths. (b) Penetration depth on one side of the DPS cladding with various core widths. (c) Simulated near-field distributions of $|E_y|$ for $f = 0.7$ GHz when the signal is input into the left end of the waveguide. The white and orange arrows show the directions of the energy flux and wave vector, respectively.

To better understand the exotic behavior of the hyperbolic guided modes, we do a full-wave simulation of the corresponding energy flux distribution. When a signal is input into the circuit-based hyperbolic waveguide, energy propagates toward the output port (i.e., the group velocity is positive), as shown in Fig. 4 (a). Two discrete ports are marked by the red dots. Figure 4(b) shows an expanded view of the distribution of Poynting vectors, which shows the enlarged energy flow in $2\times3$ units of the waveguide. These results clearly show that the Poynting vectors are strongly confined within the waveguide and decay quickly outside the waveguide. The HMM core is colored red to facilitate viewing. One of the characteristics of hyperbolic guided mode based on circuit-based HMM is the anti-parallelism for group and phase velocities, contrasted to the parallelism in DPS material. Therefore, the requirement of wave vector conservation in the interface leads to the direction of Poynting vector in HMM is antiparallel to that in DPS medium, thus forming vortex interface mode. This physical mechanism is similar to the double-negative metamaterials [72, 73].

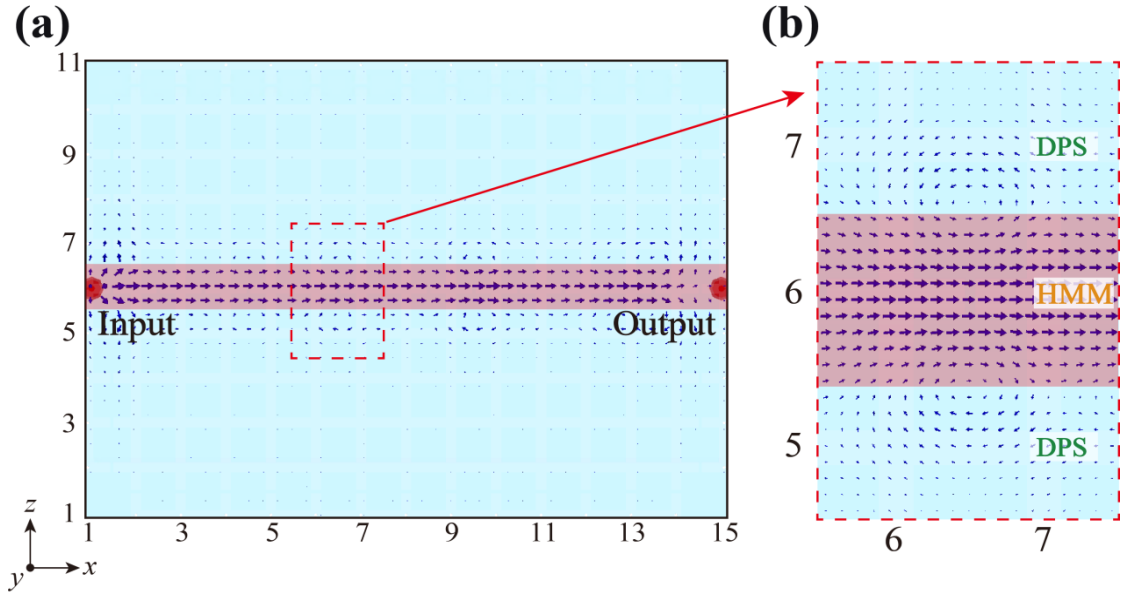

FIG. 4. (a) Distribution of Poynting vectors of guided mode in hyperbolic waveguide. Input and output ports are marked by the red dots. (b) Enlarged view of region enclosed by dashed red rectangle in panel (a). The HMM core is colored red to facilitate viewing.

To directly determine that the phase velocity and group velocity are oriented in opposite directions in the circuit-based hyperbolic waveguide, we simulate the electric-field distribution for different frequencies and phases. The width $d = 12$ mm of the HMM core remains fixed, and the frequency $f = 0.7$ GHz. Figure 5(a) shows the distribution of the electric field $E_y$ for four different phases $\varphi = 0°$, $60°$, $120°$, and $180°$. To determine the direction of the phase velocity, we choose a specific position where the sign of field strength is positive, as shown by the dotted white circles in Fig. 5. The reference line is the white dashed line. Figure 5(a) shows that, with increasing phase, the white reference circle moves toward the input port (i.e., the phase velocity is negative). For other frequencies of the hyperbolic guided modes, such as 0.75 and 0.8 GHz, the phase characteristics are similar, as shown in Figs. 5(b) and 5(c), respectively. Figures 4 and 5 demonstrate the negative phase velocity and positive group velocity in circuit-based hyperbolic waveguides. In addition, the confinement strength of the different guided modes can be determined from the evanescent fields that extend outside the hyperbolic waveguide in Fig. 5. The guided modes with lower frequency are more strongly confined, which is consistent with the spectrum of penetration depth shown in Fig. 2(b).

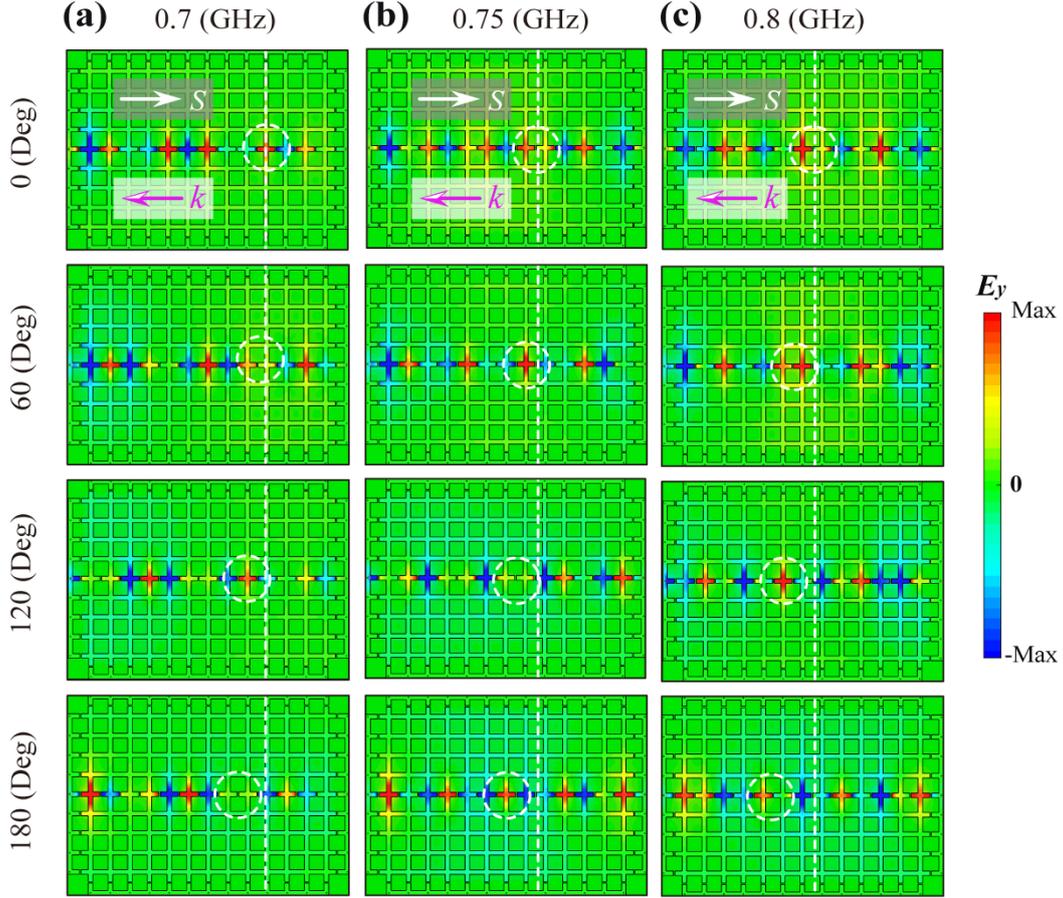

FIG. 5. Simulated near-field distributions of $E_y$ in hyperbolic waveguide for three different frequencies: (a) 0.7 GHz, (b) 0.75 GHz, and (c) 0.8 GHz with $\varphi = 0°$, $60°$, $120°$, and $180°$, respectively. The reference positions are marked by the white dashed lines.

Given that waveguide bending is an important factor in practical applications, we study herein how three modifications of the interface affect the transmission of electromagnetic waves. Specifically, we introduce a 90° turn, two 90° turns, and widen the core of the waveguide, as shown schematically in Figs. 6(a), 6(c), and 6(e), respectively. The corresponding distributions of the electric field $|E_y|$ are shown in Figs. 6(b), 6(d), and 6(f), respectively. The simulated field patterns show that, for these modifications of the interface, the hyperbolic guided mode is almost unaffected and continues to maintain efficient transmission. This indicates that guided modes in the circuit-based waveguide are robust against these interface modifications, which may be very useful in practical applications.

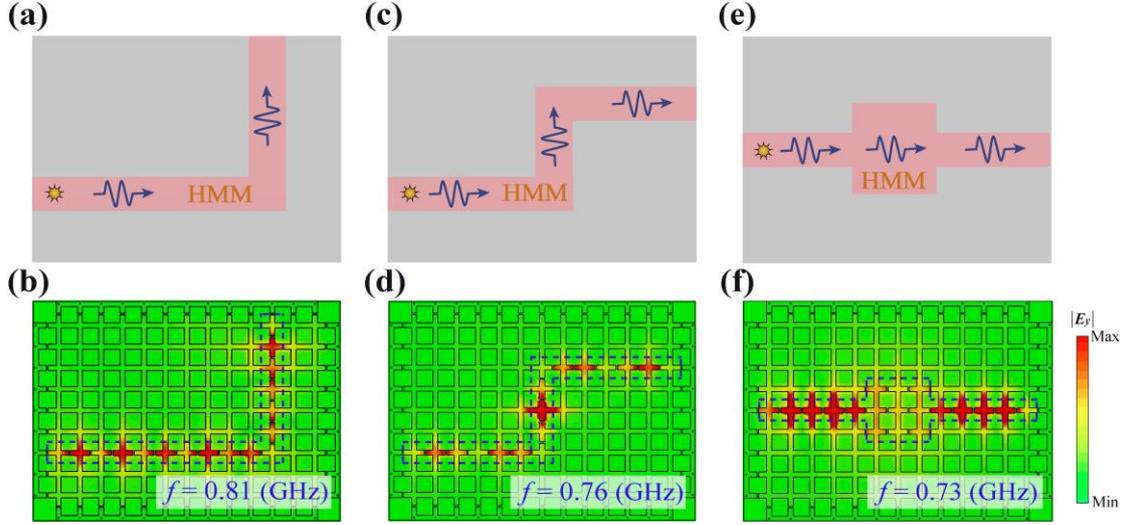

FIG. 6. (a) Schematic of hyperbolic waveguide with a 90° turn. (b) Corresponding near-field distributions of $|E_y|$ at $f = 0.81$ GHz. (c) Schematic of hyperbolic waveguide with two 90° turns. (d) Corresponding near-field distributions of $|E_y|$ at $f = 0.76$ GHz. (e) Schematic of hyperbolic waveguide with widened central section. (f) Corresponding near-field distributions of $|E_y|$ at $f = 0.73$ GHz. The HMM core is marked by the white dashed outlines.

## III. ABNORMAL PHOTONIC SPIN HALL EFFECT AND ULTRACOMPACT BACKWARD COUPLER BASED ON HYPERBOLIC WAVEGUIDES

Photons with different circular polarizations (optical spin) may propagate in different directions, which is referred to as the PSHE [70, 74]. Optical spin-orbit locking of the PSHE has attracted significant research attention and may be useful in many interesting fields of physics, such as chiral quantum optics [75] and topological photonics [76] in the near-field regime. In addition, the unidirectional excitation of the optical modes is a fundamental prerequisite for numerous photonic applications, such as polarization beam splitters and directional radiation antennas. To date, the PSHE associated with high-$k$ bulk modes in HMMs has been demonstrated [70, 77]. Here, we study abnormal directional excitation in a circuit-based hyperbolic waveguide. A near-field spin source couples only with one guided mode in a specific propagating direction determined by the rotation direction of the source [78]. For the normal PSHE, leftward (rightward) unidirectional excitation occurs for a counterclockwise (clockwise) near-field spin source [70]. The guided mode with directional wave vector in PSHE is come from the spin-momentum locking mechanism. And the propagating direction is determined by the parallelism or anti-parallelism for the wave vector and energy flow. For normal waveguide, the direction of directional transmission is the same as that of wave vector because of the same direction of wave vector and energy flow. However, for hyperbolic waveguide, the direction of directional transmission is reversed because the directions of momentum and energy flow of waveguide mode are opposite. Figure 7(a) shows a schematic illustration of the abnormal PSHE in a hyperbolic waveguide (See the dynamic properties of the waveguide modes in the supplementary materials). The simulated electric-field distributions in Figs. 7(c) and 7(e) show that, for the counterclockwise (clockwise) near-field spin source, the guided mode along the interface runs from the left (right) to the right (left). Therefore, the directional excitation in the hyperbolic waveguide is opposite that of the normal PSHE. For comparison, we also study the directional excitation in the circuit-based normal waveguide [see Fig. 7(b)] (see more details of the circuit-based normal DPS waveguides in supplementary materials). The rightward (leftward) unidirectional excitation of counterclockwise (clockwise) near-field spin source in the hyperbolic waveguide changes to leftward (rightward) unidirectional excitation in the normal waveguide, as shown in Figs. 7(d) and 7(f). The extraordinary guided mode in HMMs makes them a good research platform to study abundant unidirectional transmission.

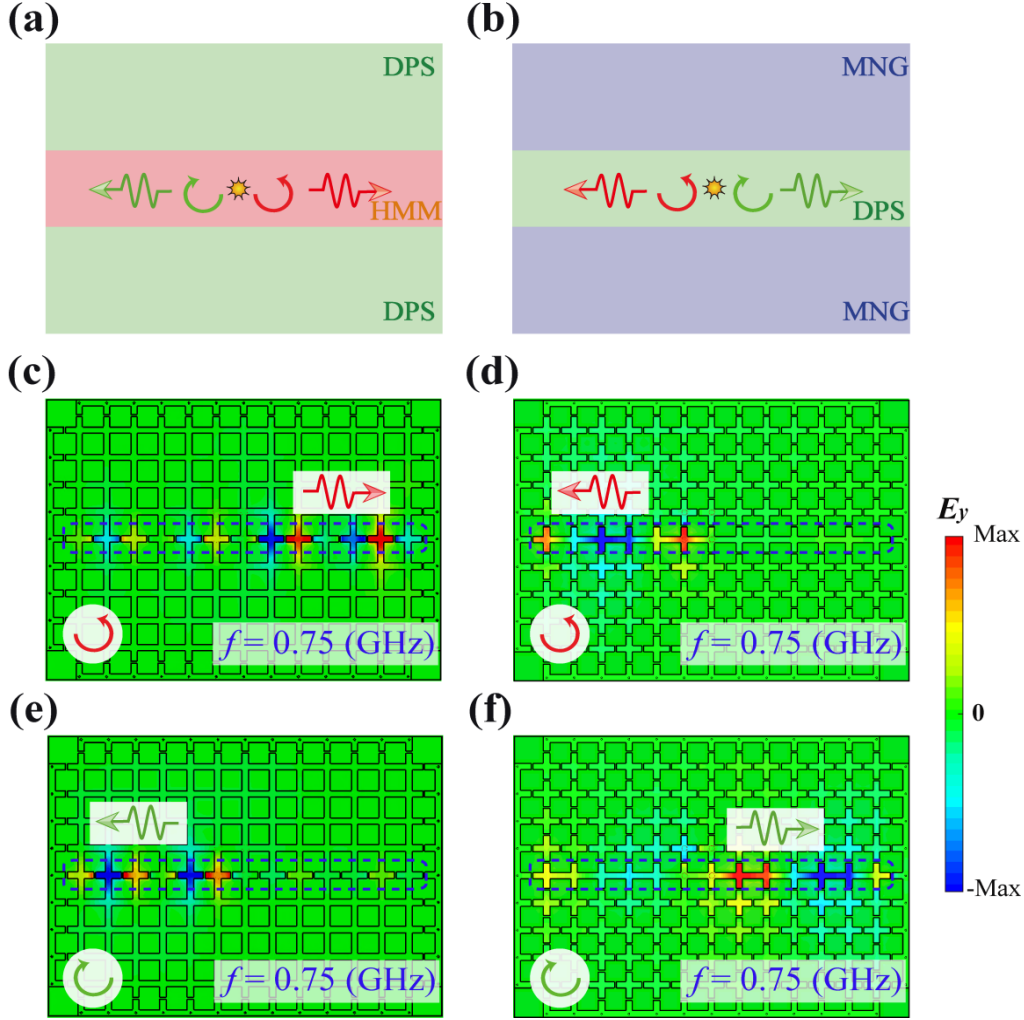

FIG. 7. (a) Schematic of the PSHE in a HMM waveguide and (b) in a DPS waveguide. The source with specific handedness excites only a single guided mode with a specific propagating direction. Anomalous unidirectional excitation occurs in the HMM waveguide. For the counterclockwise-rotating source, only the guided mode propagating from right to left and left to right will be excited in (c) the HMM waveguide and (d) in the DPS waveguide, respectively. But for the clockwise-rotating source, only the guided mode propagating from left to right and right to left will be excited in (e) the HMM waveguide and (f) in the DPS waveguide, respectively.

The abnormal dispersion of the hyperbolic waveguide mode not only allows for anomalous directional excitation but can also serve in the miniaturized backward coupler. Optical couplers have also attracted considerable attention because they may find applications in switches, routing, and power dividers [71, 72]. First, we consider the coupler composed of two hyperbolic waveguides, as shown schematically in Fig. 8(a). The conservation condition of the tangential wave vector imposes forward coupling in this coupler, and Fig. 8(d) shows the corresponding electric-field distribution. In particular, given the large effective refractive index of the hyperbolic waveguide, multiple coupling occurs within the designed finite length. Next, we consider a coupler composed of two normal waveguides, as shown schematically in Fig. 8(b). Similar to the coupler composed of two hyperbolic waveguides in Fig. 8(d), the input electromagnetic waves are coupled forward, and Fig. 8(e) shows the electric-field distribution. Given the small effective refractive index of the normal waveguide, only a single coupling takes place within the designed finite length. Finally, we consider a composite coupler, which is composed of a hyperbolic waveguide and a normal waveguide, as shown schematically shown in Fig. 8(c). As opposed to the forward coupling shown in Figs. 8(d) and 8(e), the coupling direction of the composite coupler is reversed, and Fig. 8(f) shows the corresponding electric-field distribution. The mechanism of backward coupling prevents the electromagnetic waves from coupling to return to the input waveguide after coupling once, which greatly increases the coupling efficiency. This planar subwavelength backward coupler provides a new scheme for coupler design in the microwave regime.

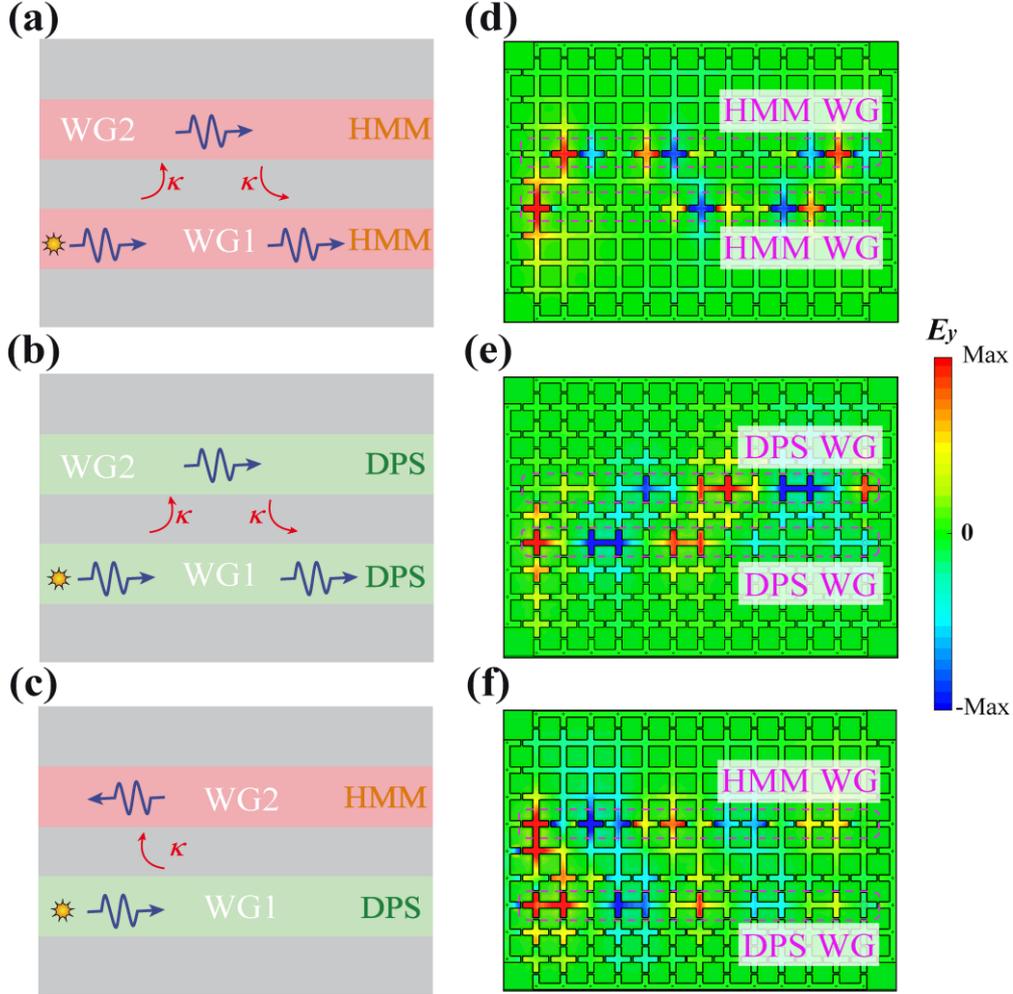

FIG. 8. (a) Schematic of forward HMM-HMM coupler, (b) forward DPS-DPS coupler, and (c) backward DPS-HMM coupler. (d) Simulated near-field distribution of $E_y$ for HMM-HMM coupler at $f = 0.85$ GHz. Similar to panel (d), but for the (e) DPS-DPS coupler and (f) DPS-HMM coupler.

To end this section, we discuss the experimental work that demonstrates the backward coupler for the circuit-based magnetic hyperbolic waveguide. Figures 9(a)-9(c) show the three samples correspond to Fig. 8(a)-8(c), respectively. In the experiment, signals were generated from a vector network analyzer (Agilent PNA Network Analyzer N5222A). A single monopole source served as a point source to excite the circuit-based prototype. A 2-mm-long homemade rod antenna served to measure the out-of-plane electric field $E_y$ at a fixed height 1 mm above the planar microstrip. The sample was placed on an automatic translation device with a scanning step of 2 mm, allowing for accurate probing of the field distribution via a near-field scanning measurement. The field amplitudes are normalized according to their respective maximum amplitude. For the couplers composed of two hyperbolic waveguides or two normal waveguides, the input electromagnetic waves are coupled forward, as shown in Figs. 9(d) and 9(e), respectively. However, figure 9(f) shows clearly that the electromagnetic wave input from DPS waveguide will be directly coupled into HMM waveguide and the energy flow will be reversed in the backward DPS-HMM coupler, avoiding the electromagnetic wave from coupling to return to the input waveguide after coupling once. The measured results [see Figs. 9(d)-9(f)] are consistent with the simulated results shown in Figs. 8(d)-8(f). These results demonstrate the planar miniaturized backward coupler based on the circuit-based hyperbolic waveguide. Overall, these results not only validate the design of the magnetic hyperbolic waveguide and reveal the abnormal dispersion relation but also demonstrate a novel miniaturized backward coupler built from the circuit-based hyperbolic waveguide. Specially, this backward coupler based on circuit-based HMMs can also be used to design novel photonic topological insulators [79].

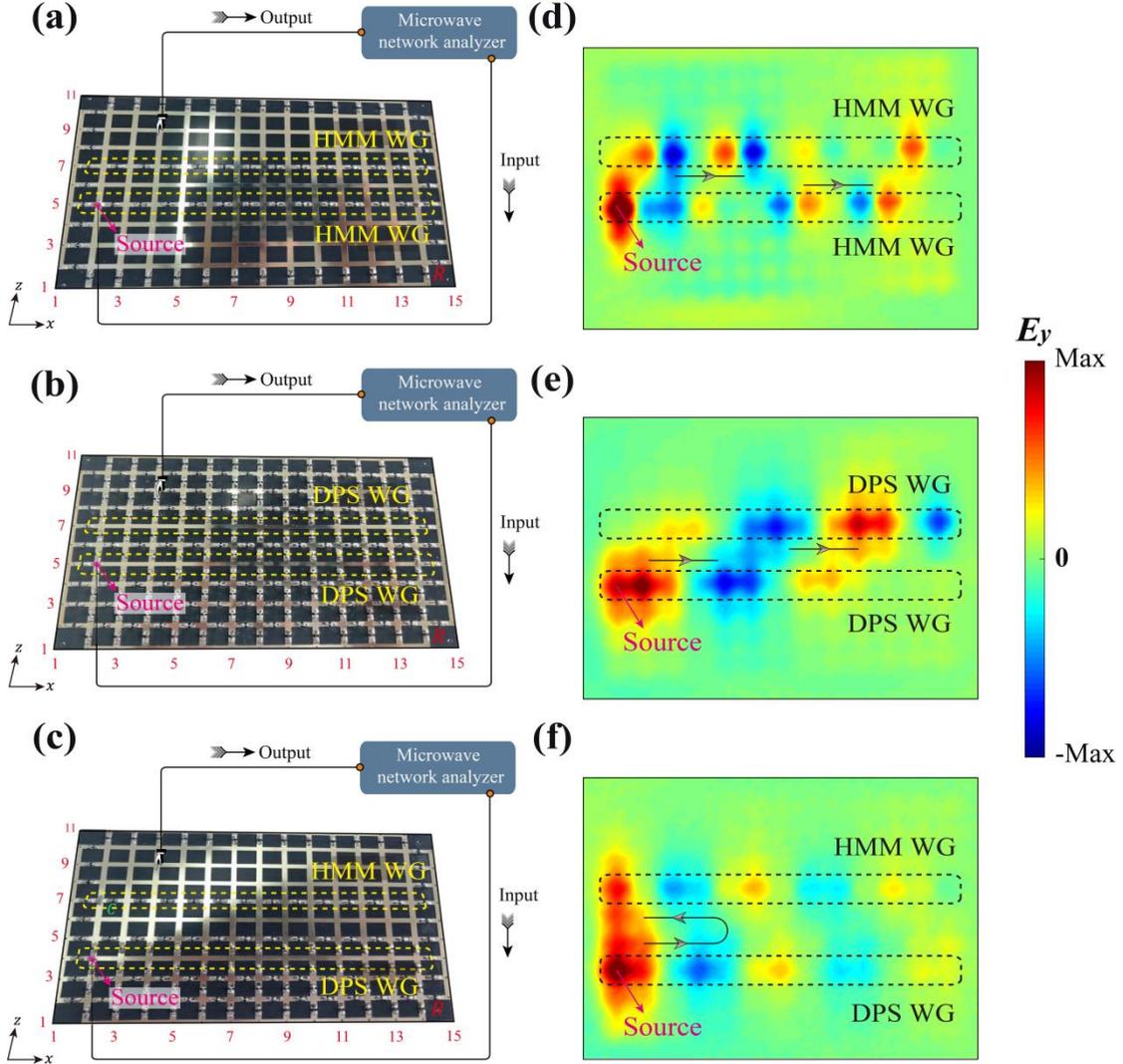

FIG. 9. (a) Schematic of experiment involving TL-based forward HMM-HMM coupler, (b) forward DPS-DPS coupler, and (c) backward DPS-HMM coupler. The prototypes of the 2D TL has $15 \times 11$ unit cells, where the waveguides are marked by the yellow dashed lines. (b) Normalized $E_y$ spectra measured in HMM-HMM coupler at 0.85 GHz. Similar to panel (d), but for the (e) DPS-DPS coupler and (f) DPS-HMM coupler.

## IV. CONCLUSION

In summary, we study herein the abnormal dispersion of circuit-based magnetic hyperbolic guided modes. As opposed to normal waveguides, the group velocity and phase velocity in hyperbolic waveguides are oriented in opposite directions. Based on this property, we reveal the optical abnormal spin Hall effect and design and experiment with a backward coupler based on the hyperbolic waveguide. The experimental results demonstrate the effectiveness of the hyperbolic waveguide in the microwave regime, and this design can be easily extended to a higher frequency range by using a natural 2D hyperbolic medium. Therefore, the results not only reveal the sophisticated electromagnetic functionalities of hyperbolic waveguides in the near field but also may provide opportunities for developing integrated optical devices, such as splitters, directional radiation antennas, planar miniaturized couplers.

**ACKNOWLEDGMENTS**

This work was supported by the National Key R&D Program of China (Grant No. 2016YFA0301101), the National Natural Science Foundation of China (NSFC) (Grant Nos. 11774261, 11474220, and 61621001), the Natural Science Foundation of Shanghai (Grant Nos. 17ZR1443800 and 18JC1410900), the Shanghai Super Postdoctoral Incentive Program, and the China Postdoctoral Science Foundation (Grant Nos. 2019TQ0232 and 2019M661605).